\documentclass{ws-procs9x6}            

\usepackage{CJK}

\begin{document}
\begin{CJK*}{GB}{gbsn}
\title{Binding energies and pairing gaps in semi-magic nuclei obtained using
new regularized higher-order EDF generators}

\author{K. Bennaceur,$^{1,2,3}$ J. Dobaczewski,$^{2,3,4,5}$ and Y. Gao (¸ßÔ­)$^{6}$}
\address{%
$^1$Univ Lyon, Universit\'e Lyon 1, CNRS/IN2P3, IPNL, F-69622 Villeurbanne, France \\
$^2$Department of Physics, PO Box 35 (YFL), FI-40014 University of Jyv{\"a}skyl{\"a}, Finland \\
$^3$Helsinki Institute of Physics, P.O. Box 64, FI-00014 University of Helsinki, Finland \\
$^4$Department of Physics, University of York, Heslington, York YO10 5DD, United Kingdom \\
$^5$Institute of Theoretical Physics, Faculty of Physics, University of Warsaw, ul. Pasteura 5, PL-02-093 Warsaw, Poland \\
$^6$State Key Laboratory of Nuclear Physics and Technology, School of Physics, Peking University, Beijing 100871, China
}

\begin{abstract}
We present results of the Hartree-Fock-Bogolyubov calculations
performed using nuclear energy density functionals based on
regularized functional generators at next-to-leading and
next-to-next-to-leading order. We discuss properties of binding
energies and pairing gaps determined in semi-magic spherical nuclei.
The results are compared with benchmark calculations performed for
the functional generator SLyMR0 and functional UNEDF0.
\end{abstract}

\keywords{nuclear energy density functionals, regularized functional
generators, nuclear binding energies, pairing gaps}

\end{CJK*}

\bodymatter

\section{Introduction}\label{sec:Introduction}
A quest for energy density functionals (EDFs) that would precisely
and accurately describe multitude of low-energy nuclear properties is
at present one of the most important research avenues in nuclear
physics. The problem has been addressed in quite a number of recent
studies,\cite{(Car08),(Zal08),(Rai11),(Dob12),(Rai14),(Ben14a),(Sad13),(Sad13b),(Dav15),(Ben17)}
where various extensions of the standard EDFs, used for the last 60-odd
years, were proposed.

In this conference communication, we present results of calculations
performed for semi-magic nuclei across the mass chart, using the newly
developed EDFs based on the regularized higher-order generators with
pairing.\cite{(Ben17)} As discussed in Ref.\citenum{(Ben17)}, the proposed
new parametrizations at next-to-leading order (NLO) REG2c.161026 and
next-to-next-to-leading (N$^2$LO) REG4c.161026 correspond to a
fairly low effective mass and overestimate pairing strength.
Therefore, they are not good enough to embark for them on massive
mass-table calculations. However, inexpensive calculations performed
for spherical semi-magic nuclei, which are reported on in this paper,
can constitute a useful illustration of the overall bulk properties
corresponding to the newly developed EDFs.

\section{Results}\label{sec:Results}
We performed the Hartree-Fock-Bogolyubov (HFB) calculations for all
bound semi-magic nuclei across the mass chart, using the NLO
REG2c.161026 and N$^2$LO REG4c.161026 EDFs,\cite{(Ben17)})
SLyMR0,\cite{(Sad13)}) and UNEDF0.\cite{(Kor10b)})
For UNEDF0, the Lipkin-Nogami (LN) method was used to account for
approximate particle-number restoration as in Ref.\citenum{(Sto03)}.
For the finite-range NLO and N$^2$LO generators, we solved the
non-local self-consistent equations using the newly developed code
{\sc{finres}}$_4$ (Finite-Range Self-consistent Spherical
Space-coordinate Solver),\cite{[Ben17a]} which is based on the method
proposed by Hooverman.\cite{(Hoo72)} For the zero-range generator
SLyMR0, we obtained the solutions using the spherical
solver {\sc{lenteur}},\cite{(lenteur)} and for the
quasi-local functional UNEDF0, using the code
{\sc{hosphe}},\cite{(Car10b),[Car13b]} similarly as in
Ref.\citenum{(Gao13)}.

\newlength{\mywidth}
\setlength{\mywidth}{0.91\textwidth}
\begin{figure}
\begin{center}
\begin{tabular}{c@{\hspace*{0.001\mywidth}}c}
\includegraphics[height=0.59\mywidth,angle=270,viewport=40 50 525 765]{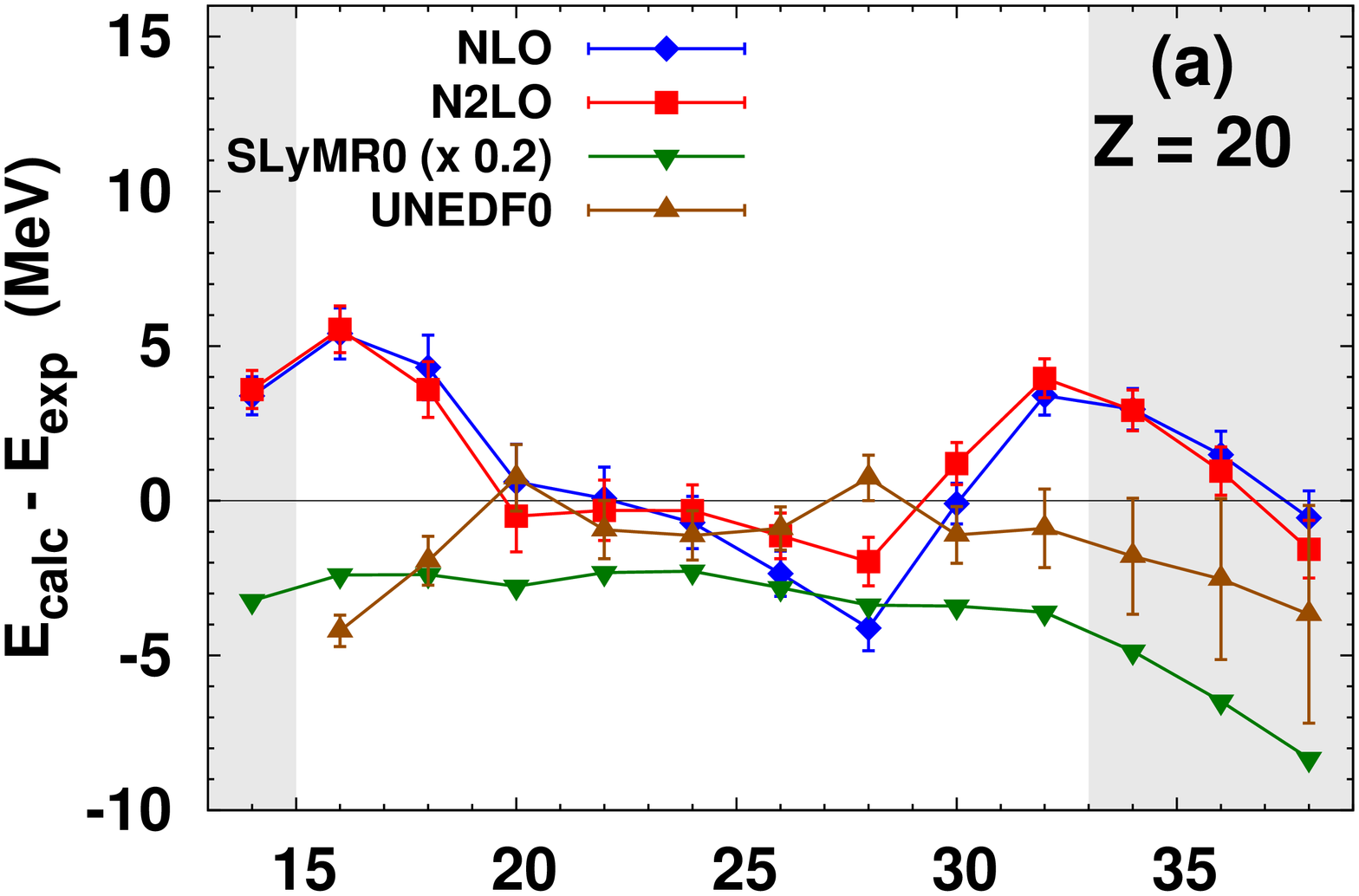} &
\includegraphics[height=0.59\mywidth,angle=270,viewport=40 175 525 890]{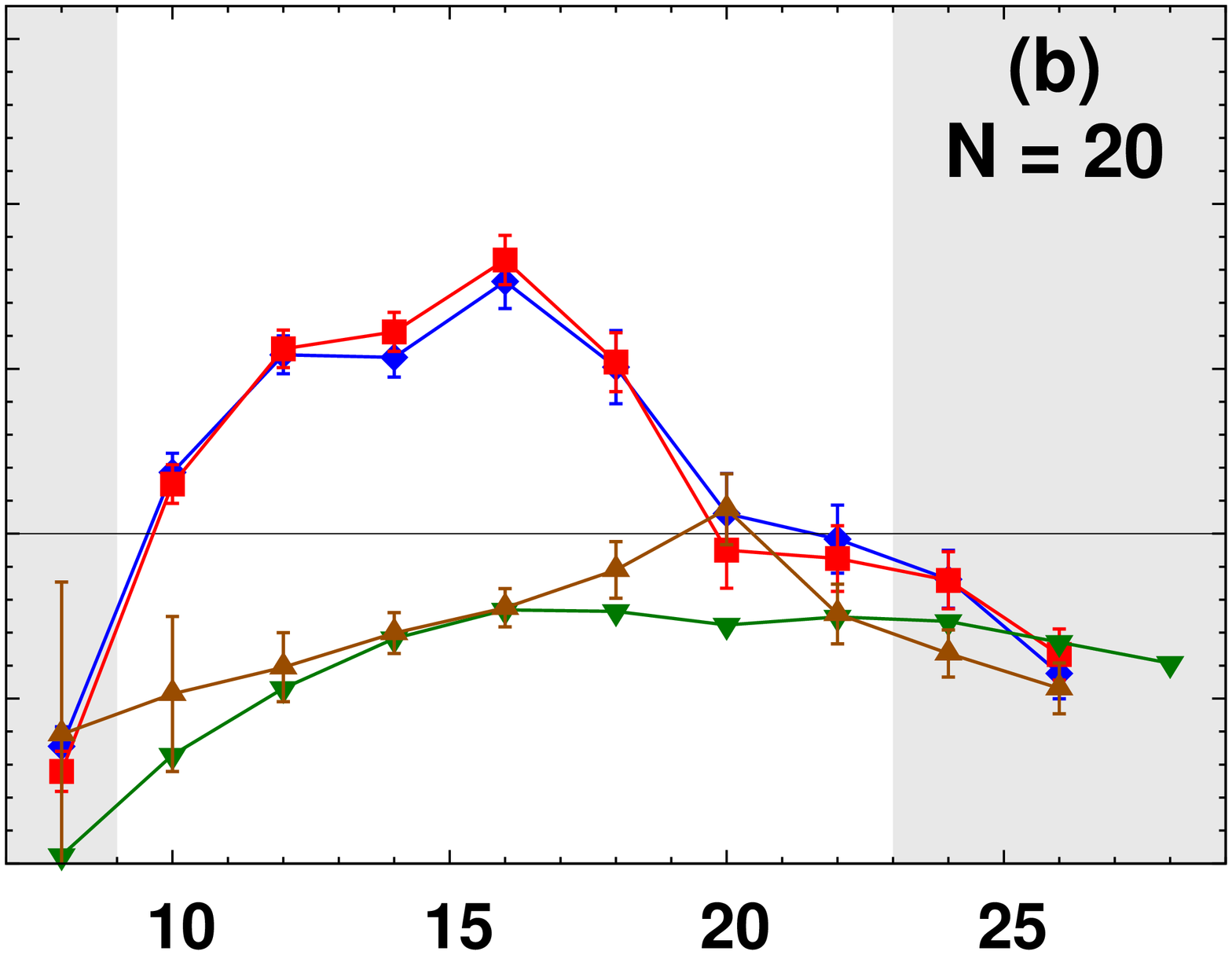} \\[-0.05\mywidth]
\includegraphics[height=0.59\mywidth,angle=270,viewport=40 50 525 765]{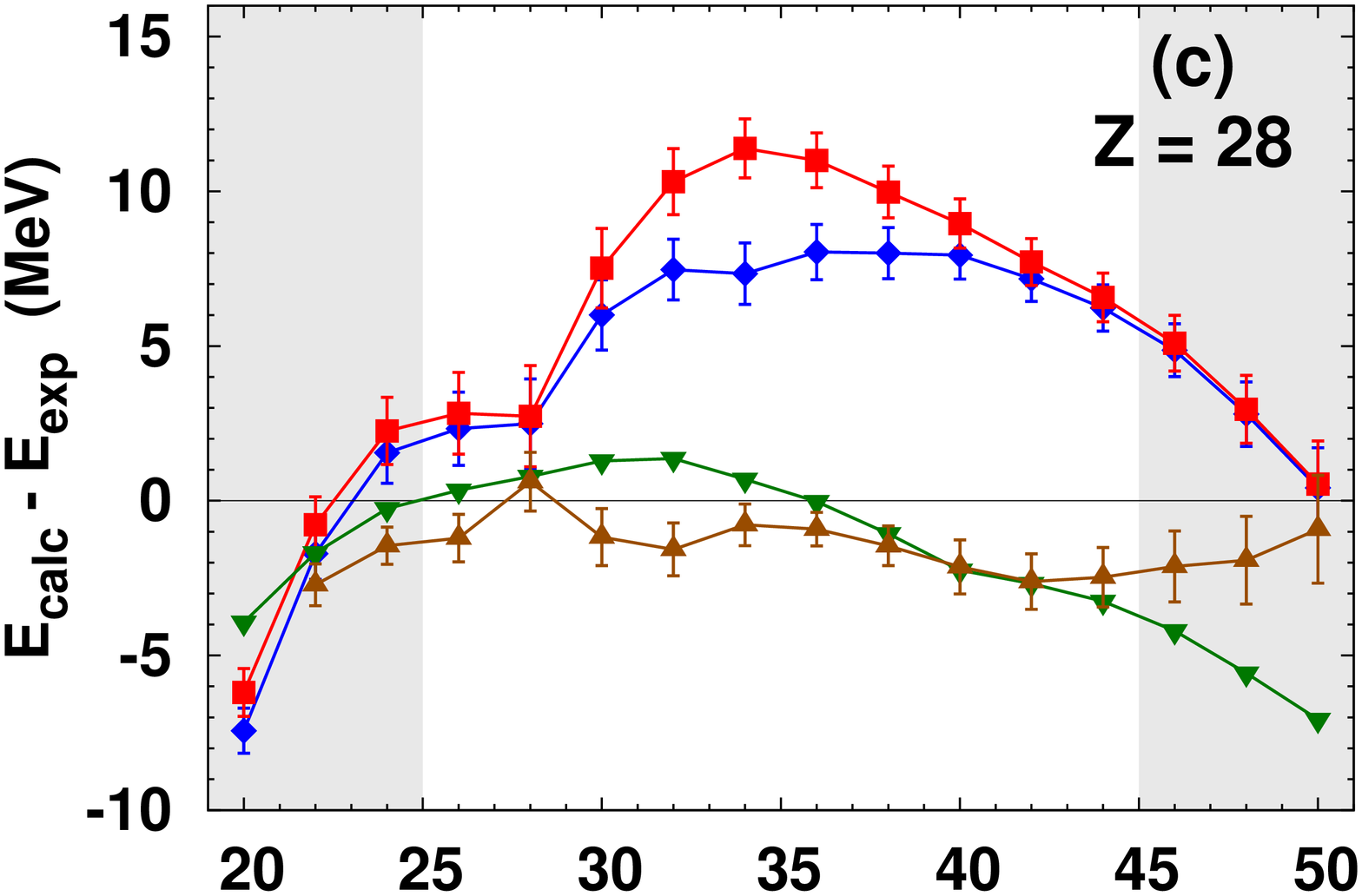} &
\includegraphics[height=0.59\mywidth,angle=270,viewport=40 175 525 890]{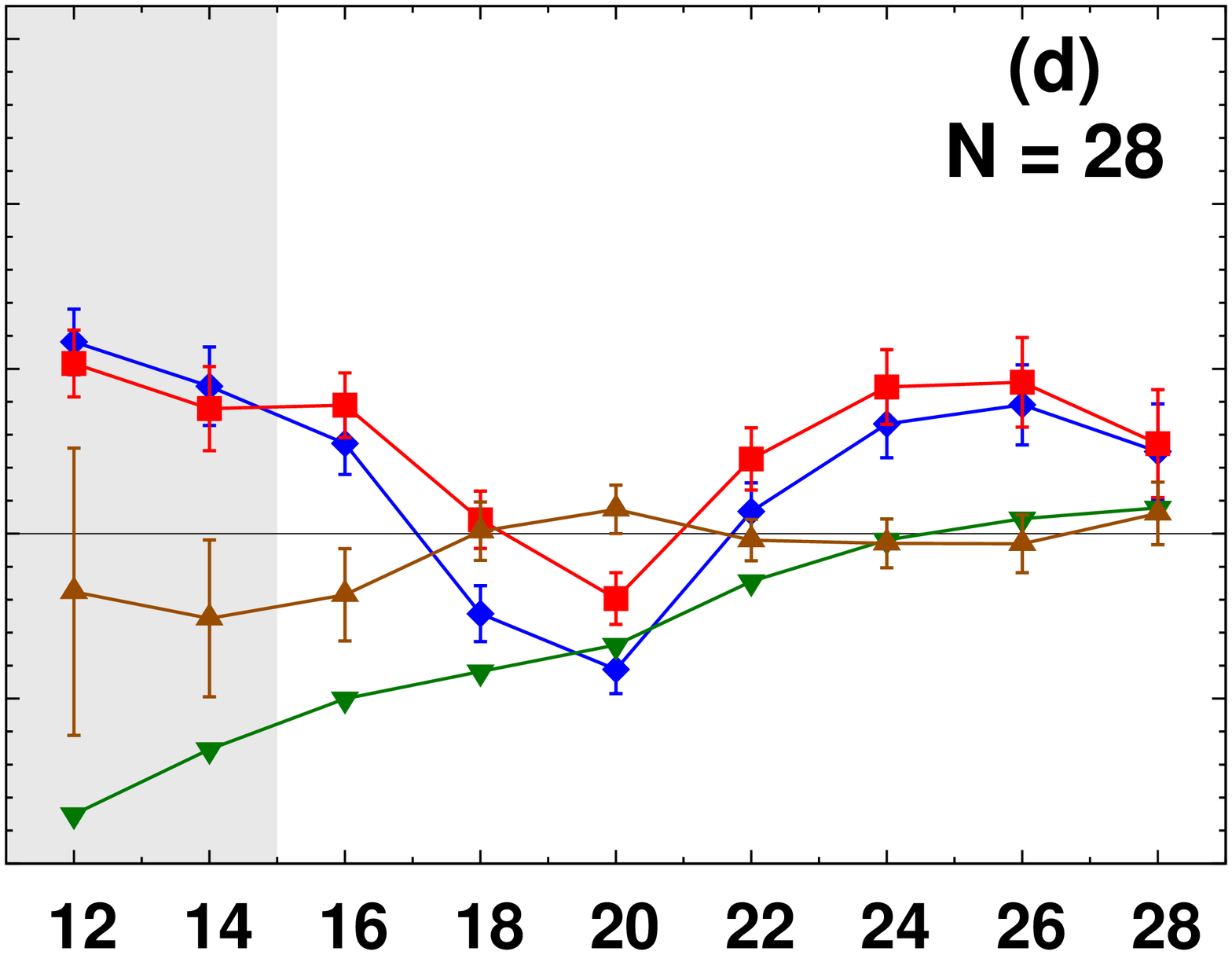} \\[-0.05\mywidth]
\includegraphics[height=0.59\mywidth,angle=270,viewport=40 50 525 765]{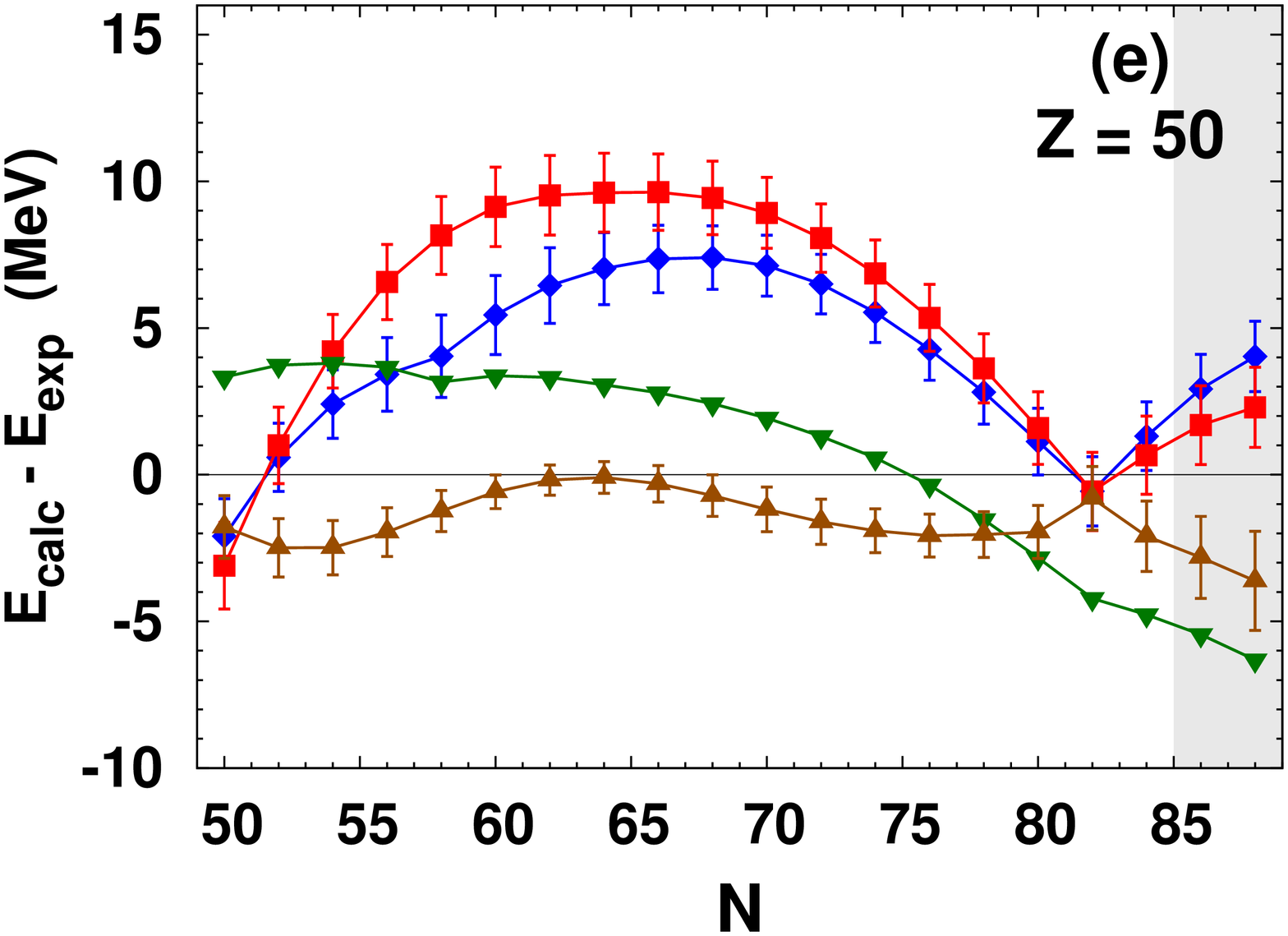} &
\includegraphics[height=0.59\mywidth,angle=270,viewport=40 175 525 890]{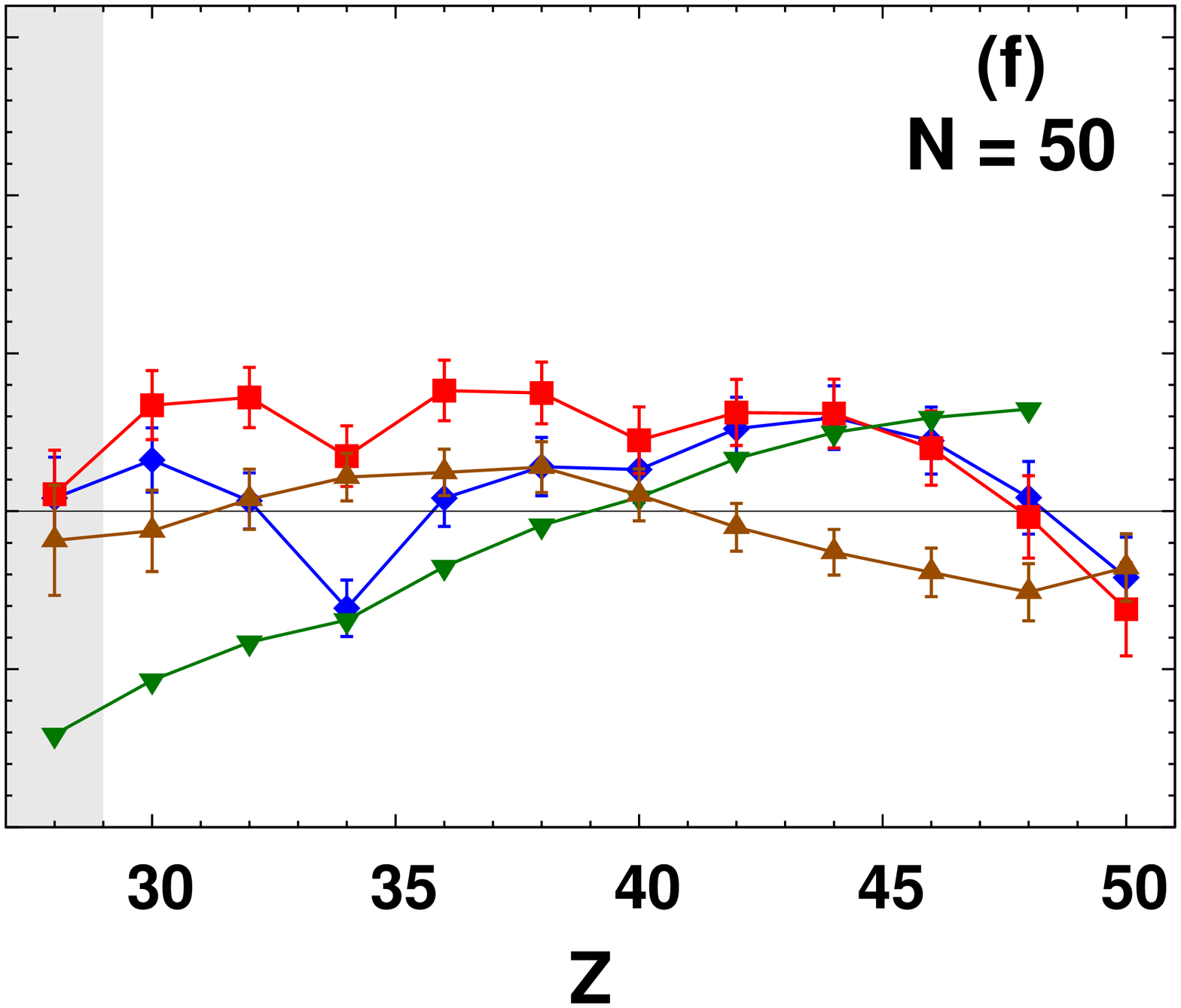}
\end{tabular}
\end{center}
\caption{Differences between calculated (HFB results) and
experimental (AME12\cite{(Wan12)}) ground-state energies of
proton-magic (left panels) and neutron-magic nuclei with $Z,N=20$,
28, or 50. Calculations were performed using EDFs
NLO REG2c.161026\protect\cite{(Ben17)}) (diamonds),
N$^2$LO REG4c.161026\protect\cite{(Ben17)}) (squares),
SLyMR0\protect\cite{(Sad13)}) (down triangles), and
UNEDF0\protect\cite{(Kor10b)}) (up triangles). For SLyMR0, to fit in the figure, the
differences were divided by a factor of five. Shaded zones correspond to
the AME12 masses taken from systematics.}
\label{fig:fig1}
\end{figure}

\begin{figure}
\begin{center}
\begin{tabular}{c@{\hspace*{0.001\mywidth}}c}
\includegraphics[height=0.59\mywidth,angle=270,viewport=40 50 525 765]{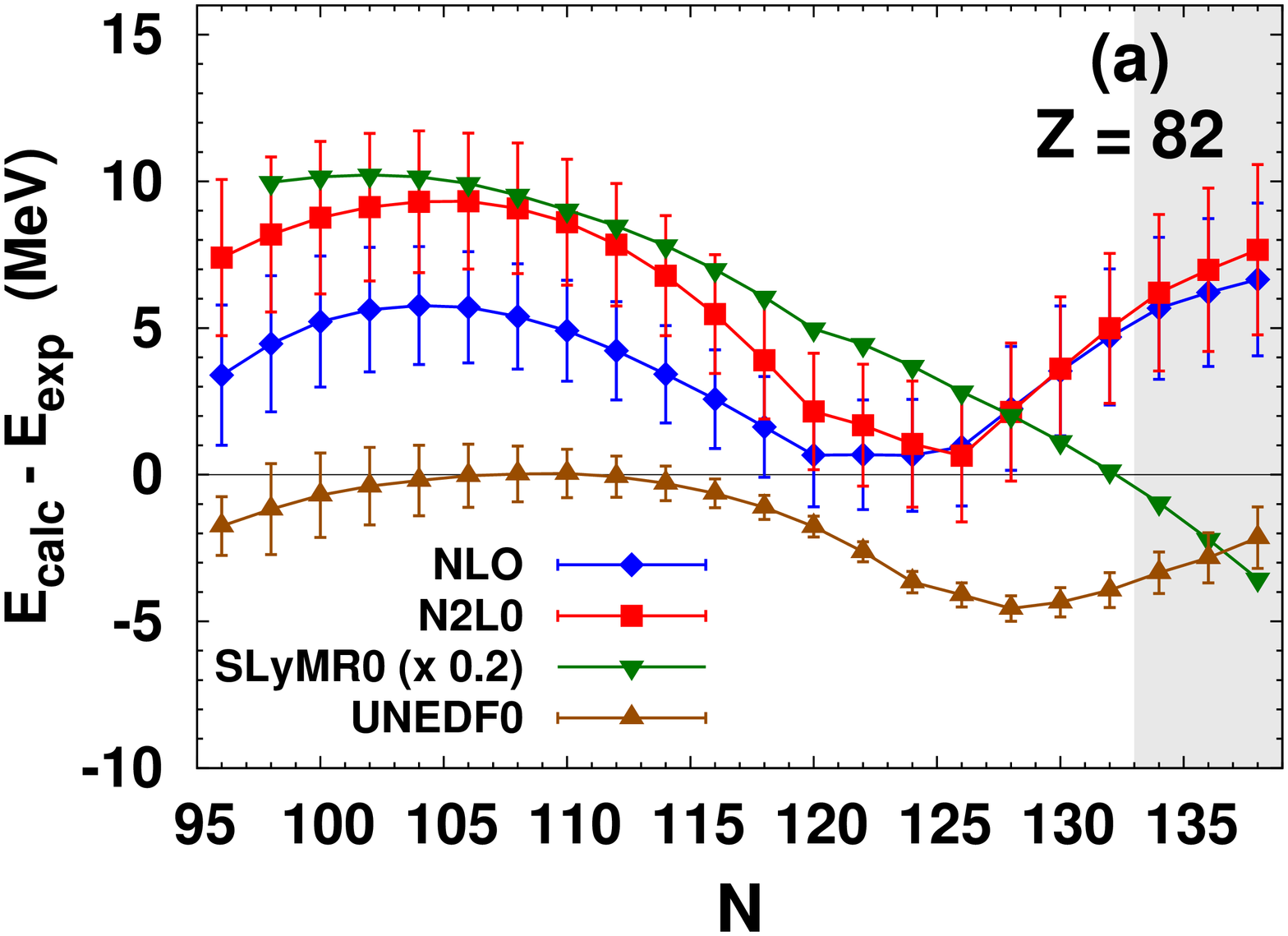} &
\includegraphics[height=0.59\mywidth,angle=270,viewport=40 175 525 890]{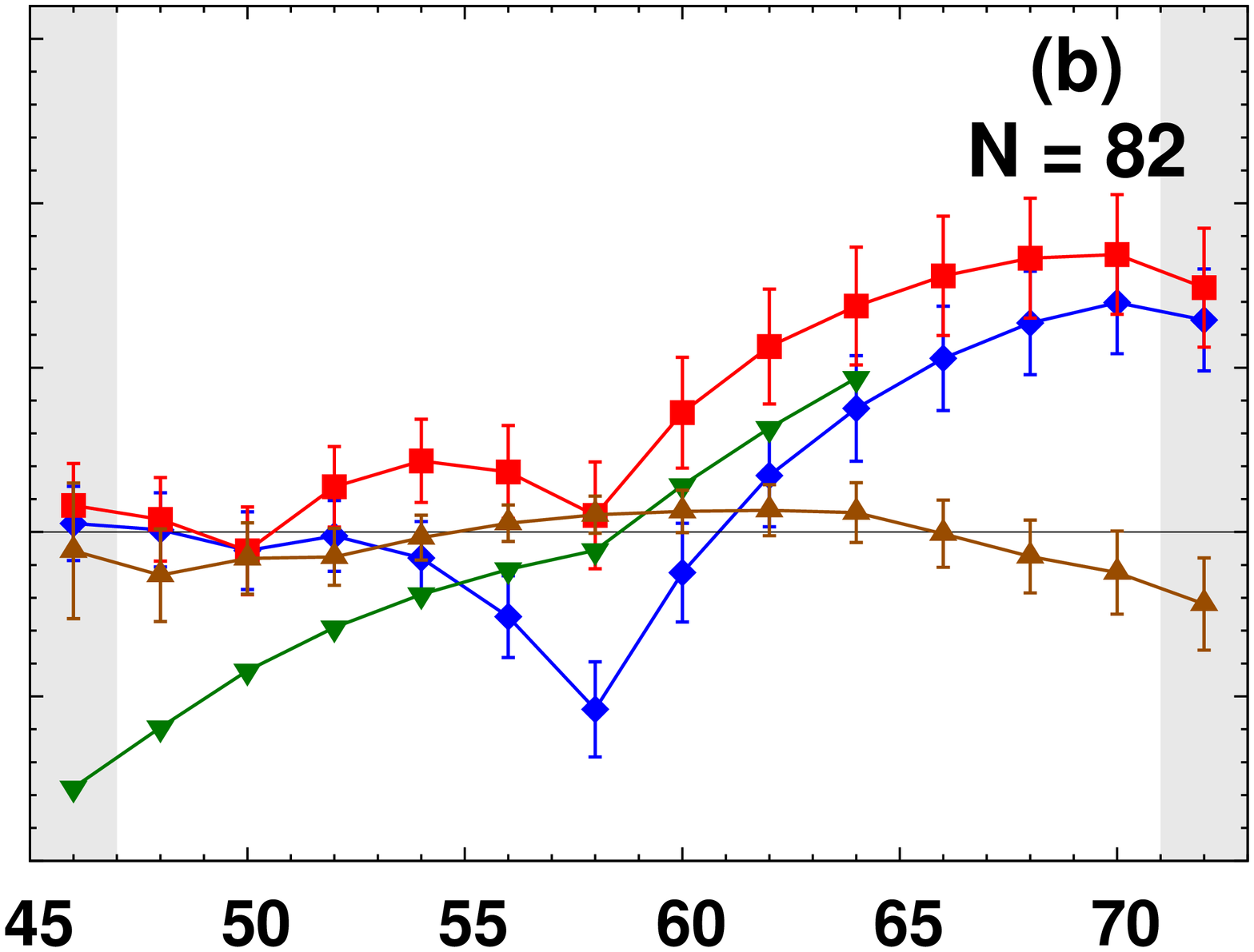} \\[-0.05\mywidth]
 &
\includegraphics[height=0.59\mywidth,angle=270,viewport=40 175 525 890]{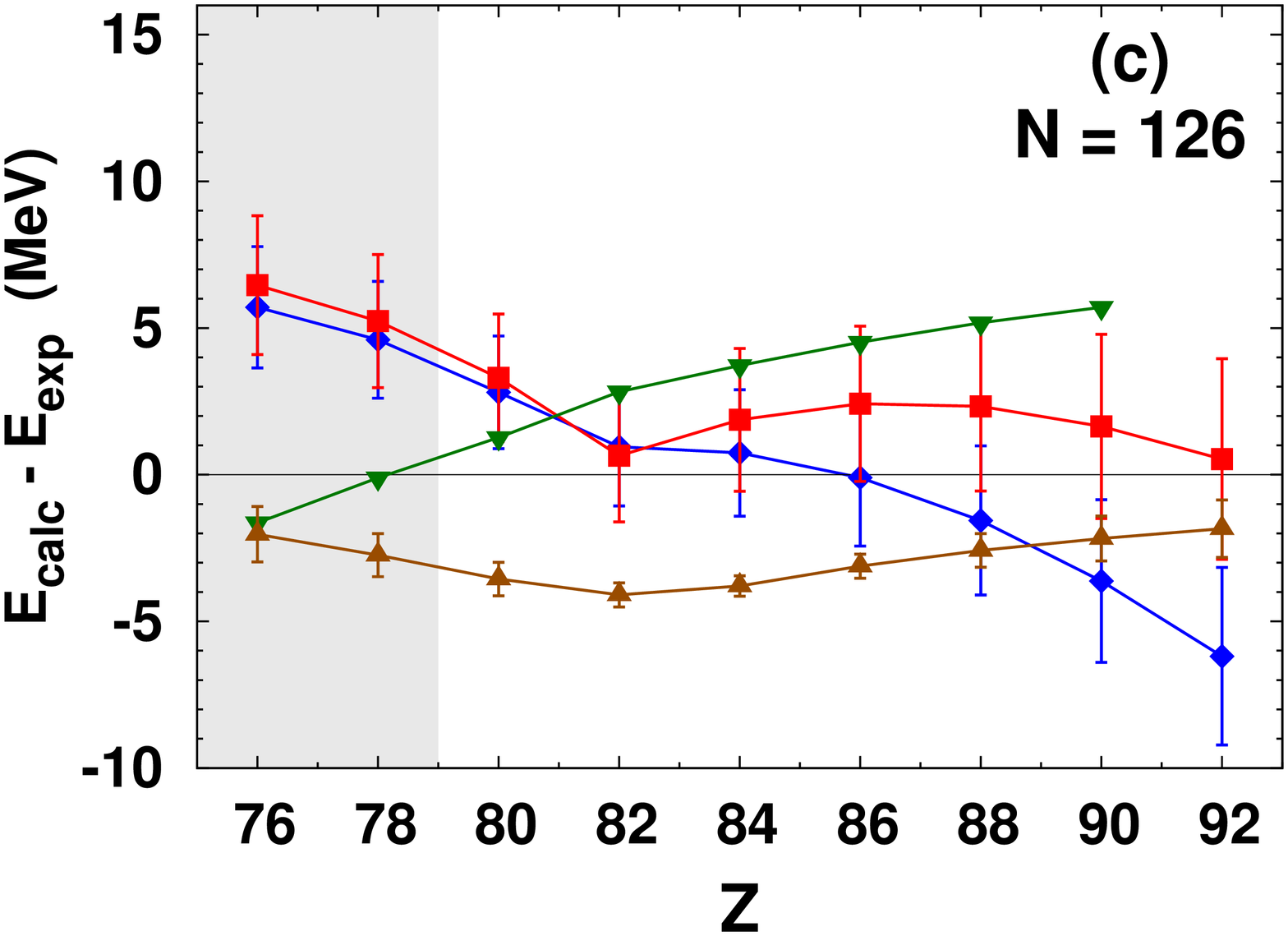}
\end{tabular}
\end{center}
\caption{Same as in Fig.~\ref{fig:fig1} but for the semi-magic nuclei with
$Z,N=82$ and $N=126$.}
\label{fig:fig2}
\end{figure}

For NLO, N$^2$LO, and UNEDF0, where the fit covariance matrices are
known, we determined statistical uncertainties of all calculated
observables according to the methodology presented in
Ref.\citenum{(Dob14)}. For SLyMR0, only the values of observables
were determined.

\begin{figure}
\begin{center}
\begin{tabular}{c@{\hspace*{0.001\mywidth}}c}
\includegraphics[height=0.59\mywidth,angle=270,viewport=40 50 525 765]{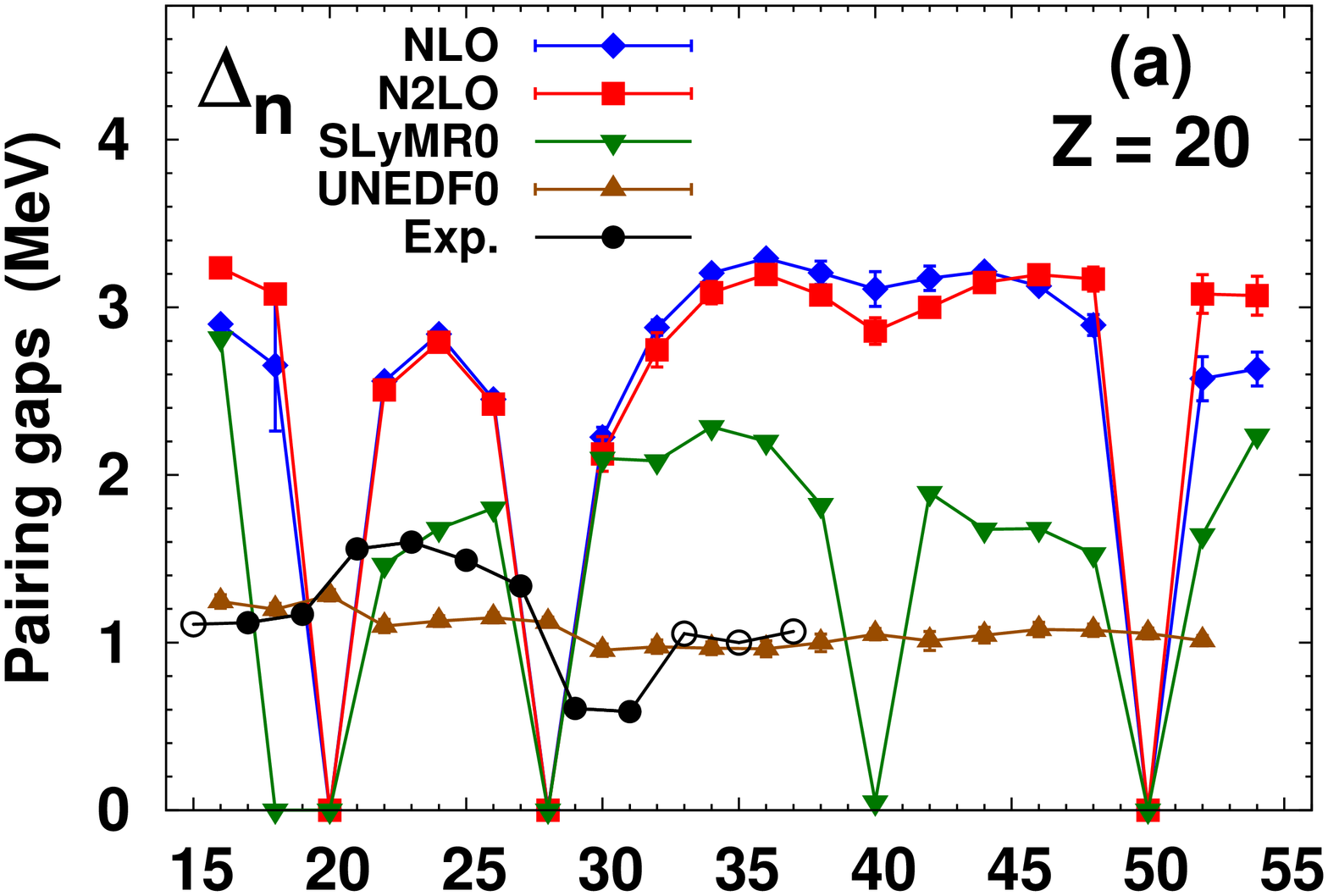} &
\includegraphics[height=0.59\mywidth,angle=270,viewport=40 175 525 890]{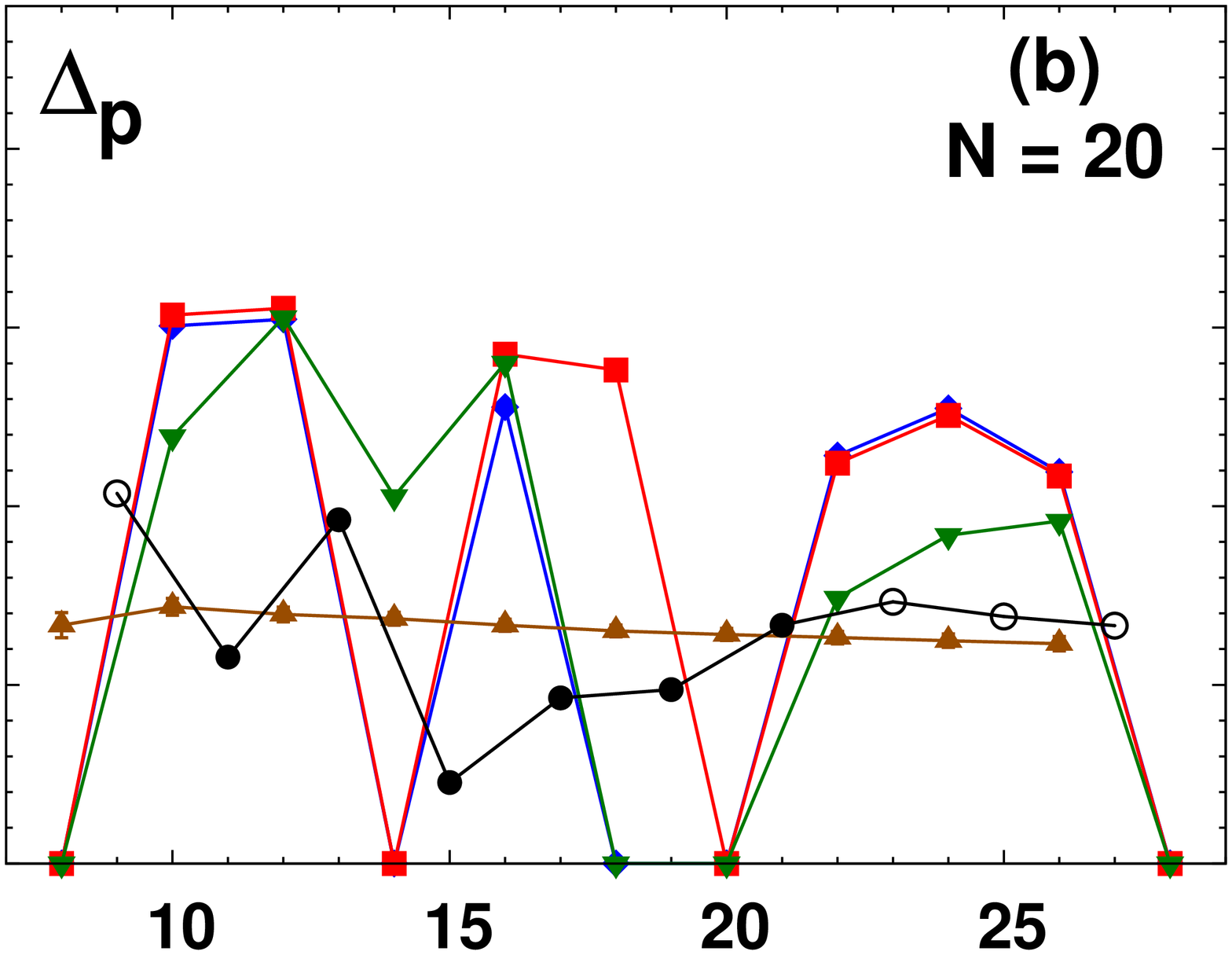} \\[-0.05\mywidth]
\includegraphics[height=0.59\mywidth,angle=270,viewport=40 50 525 765]{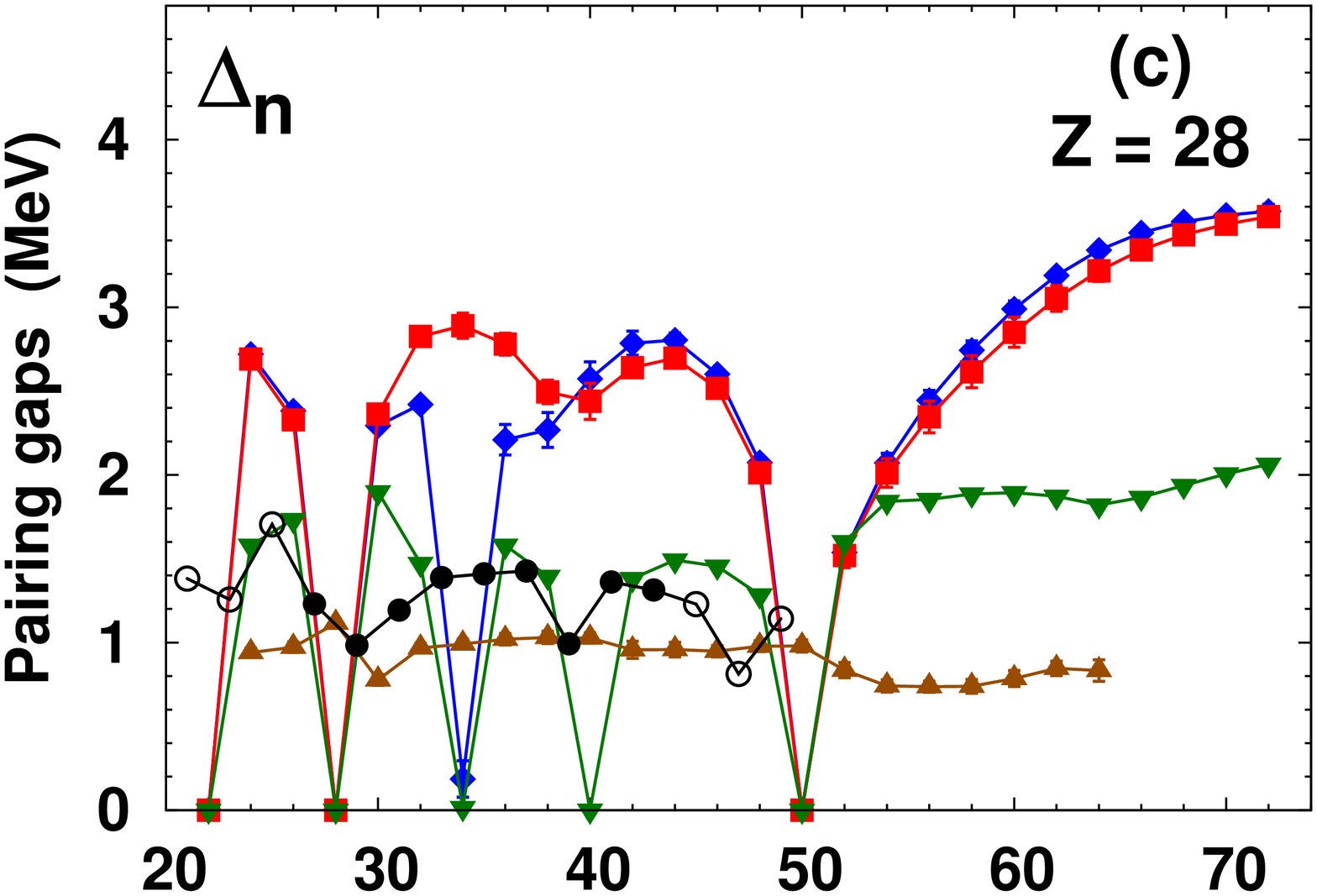} &
\includegraphics[height=0.59\mywidth,angle=270,viewport=40 175 525 890]{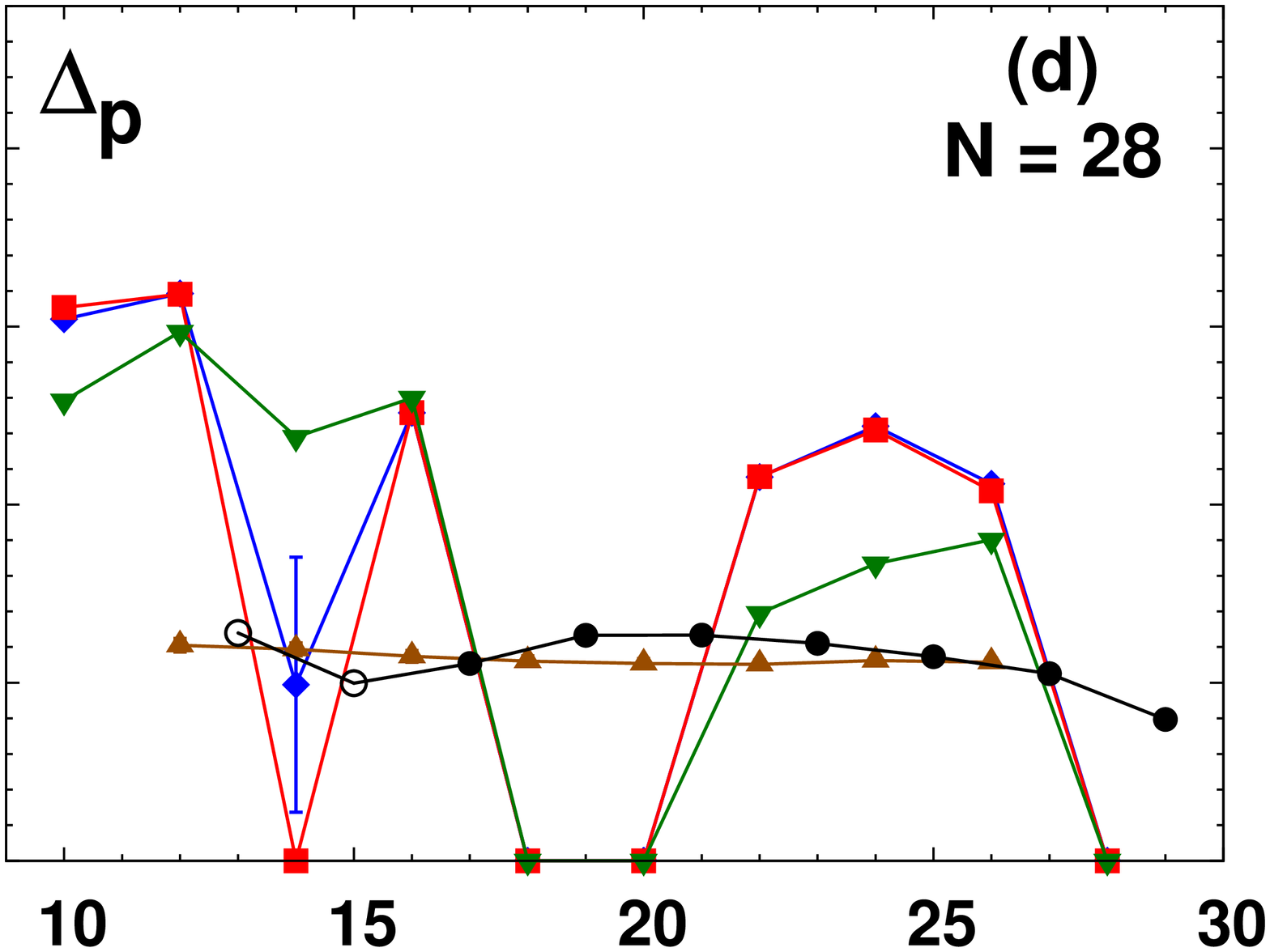} \\[-0.05\mywidth]
\includegraphics[height=0.59\mywidth,angle=270,viewport=40 50 525 765]{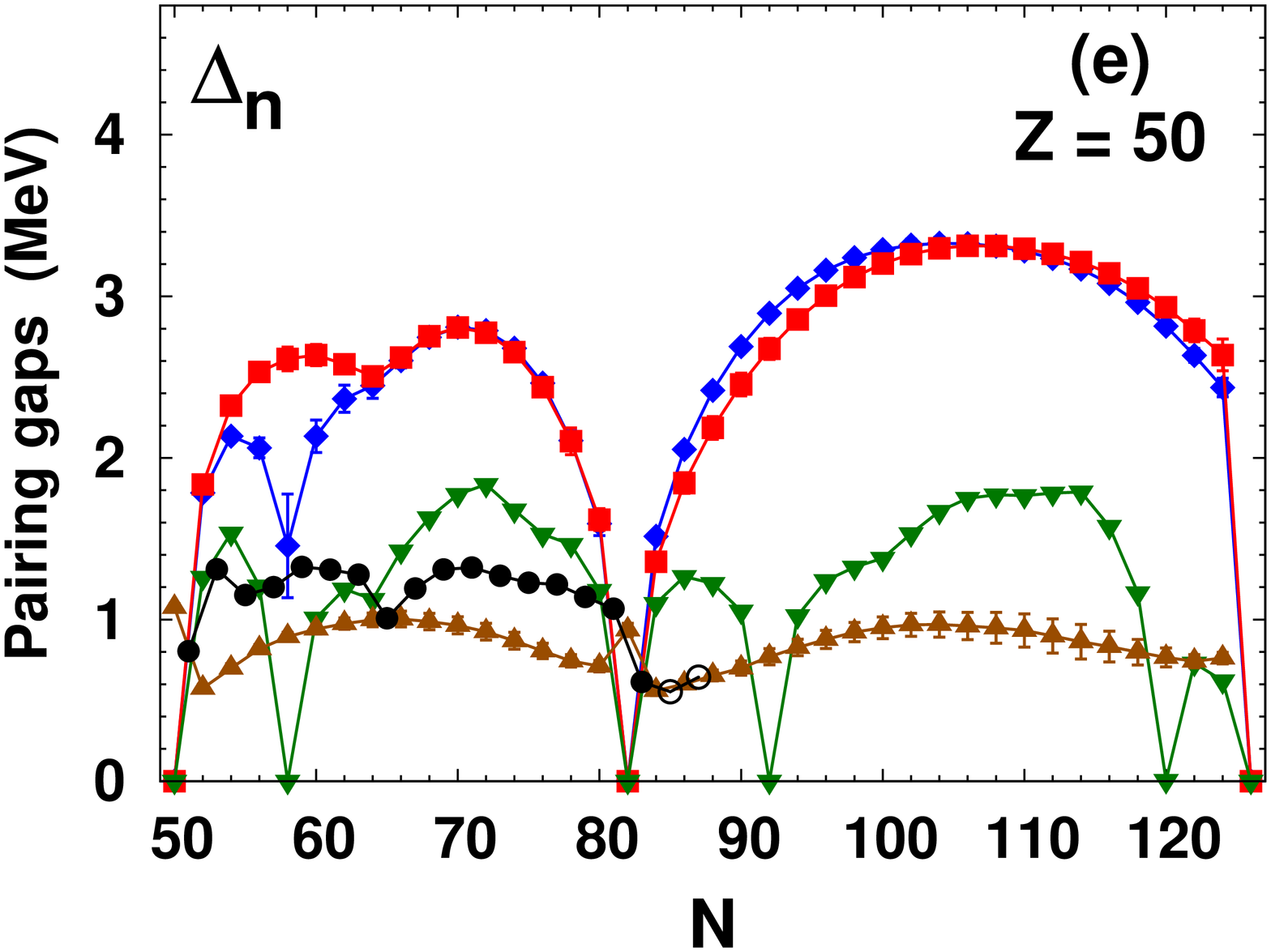} &
\includegraphics[height=0.59\mywidth,angle=270,viewport=40 175 525 890]{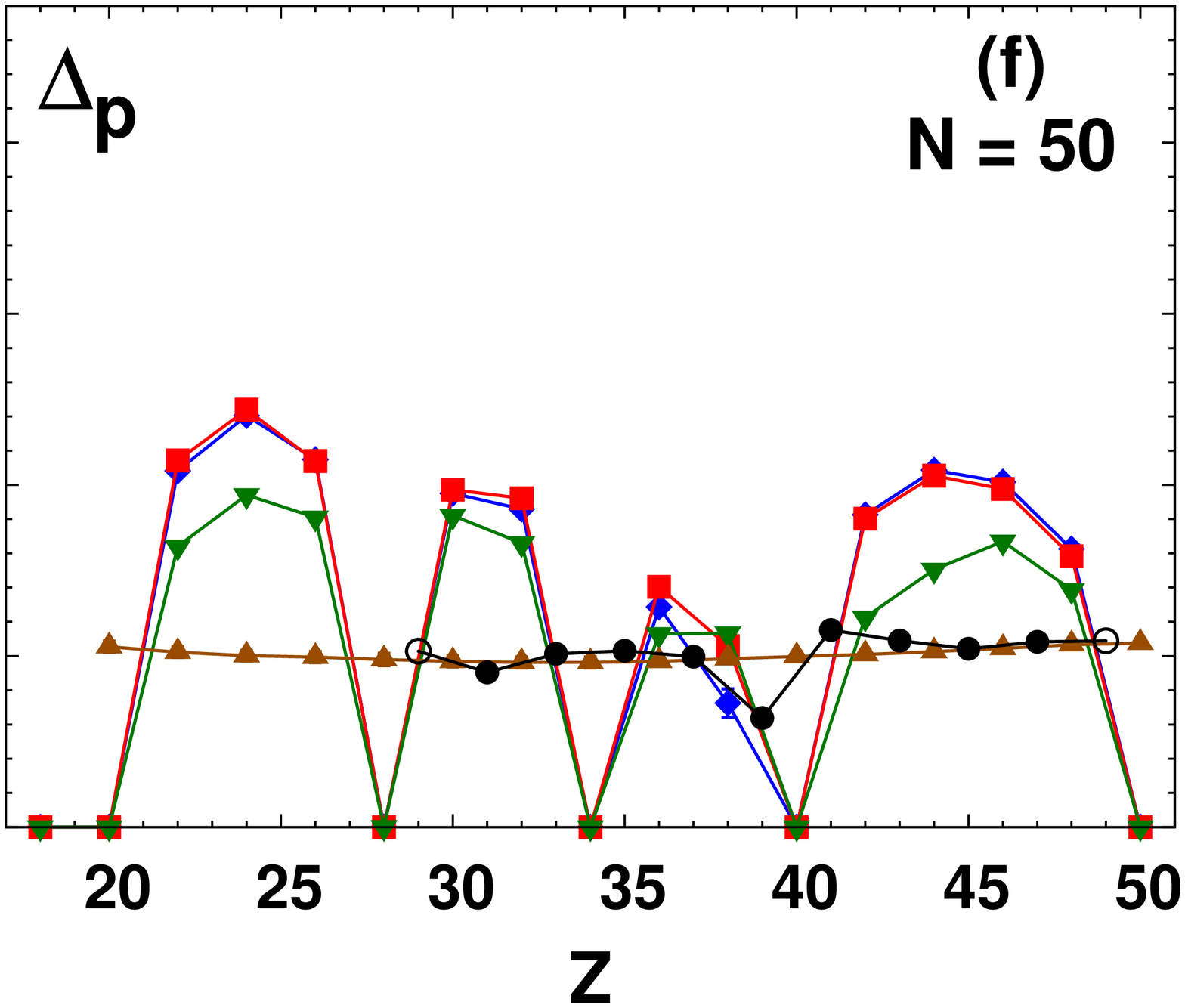}
\end{tabular}
\end{center}
\caption{Calculated (HFB results) and experimental
(AME12\cite{(Wan12)}) neutron pairing gaps of proton-magic nuclei
(left panels) and proton pairing gaps of neutron-magic nuclei with
$Z,N=20$, 28, or 50. Calculations were performed using EDFs
NLO REG2c.161026\protect\cite{(Ben17)}) (diamonds),
N$^2$LO REG4c.161026\protect\cite{(Ben17)}) (squares),
SLyMR0\protect\cite{(Sad13)}) (down triangles), and
UNEDF0\protect\cite{(Kor10b)}) (up triangles). Experimental values (circles)
correspond to the three-point mass staggering centered at odd particle
numbers.\protect\cite{(Sat98)}) Open circles indicate values, for which
at least one of the three AME12 masses were taken from systematics.}
\label{fig:fig3}
\end{figure}

In Figs.~\ref{fig:fig1} and~\ref{fig:fig2}, for nuclei where the
binding energies are known from experiment or
systematics,\cite{(Wan12)}) we show the binding-energy residuals
determined for the four studied EDFs. We note that for SLyMR0,
the obtained differences between theory and experiment are
significantly larger than those obtained for the other three EDFs, and
therefore, in the figures they were divided by a factor of five.
We also note that NLO, N$^2$LO, and SLyMR0 EDFs are
built as exact averages of two-body or many-body generators, both in
the particle-hole and pairing channels, and therefore, they are suitable
without ambiguity for beyond-mean-field and symmetry-restoration calculations.
On the other hand, EDF UNEDF0 is built as an average of a density-dependent
two-body Skyrme-type generators that are different in particle-hole
and pairing channels and with some omitted terms in the particle-hole channel.

Although the pattern of comparison with data is fairly different
among all four studied EDFs, we clearly see that the new NLO and
N$^2$LO EDFs describe data better than SLyMR0 (recall the scaling
factor of five used for SLyMR0). However, results obtained at NLO and
N$^2$LO are still fairly worse than those obtained for the standard
Skyrme-like EDF UNEDF0. In particular, for the NLO and N$^2$LO EDFs,
we see conspicuous ``arches'' of residuals between the doubly magic
nuclei. Usually these feature of calculated ground-state energies was
attributed to low effective mass, however, we note here that for
SLyMR0, which has a low effective mass too, no such an effect is
seen. We also note that the statistical uncertainties obtained for
UNEDF0 significantly increase when going towards neutron-rich
nuclei,\cite{(Gao13)} whereas those for the NLO and N$^2$LO EDFs
depend on the neutron excess much less.

\begin{figure}
\begin{center}
\begin{tabular}{c@{\hspace*{0.001\mywidth}}c}
\includegraphics[height=0.59\mywidth,angle=270,viewport=40 50 525 765]{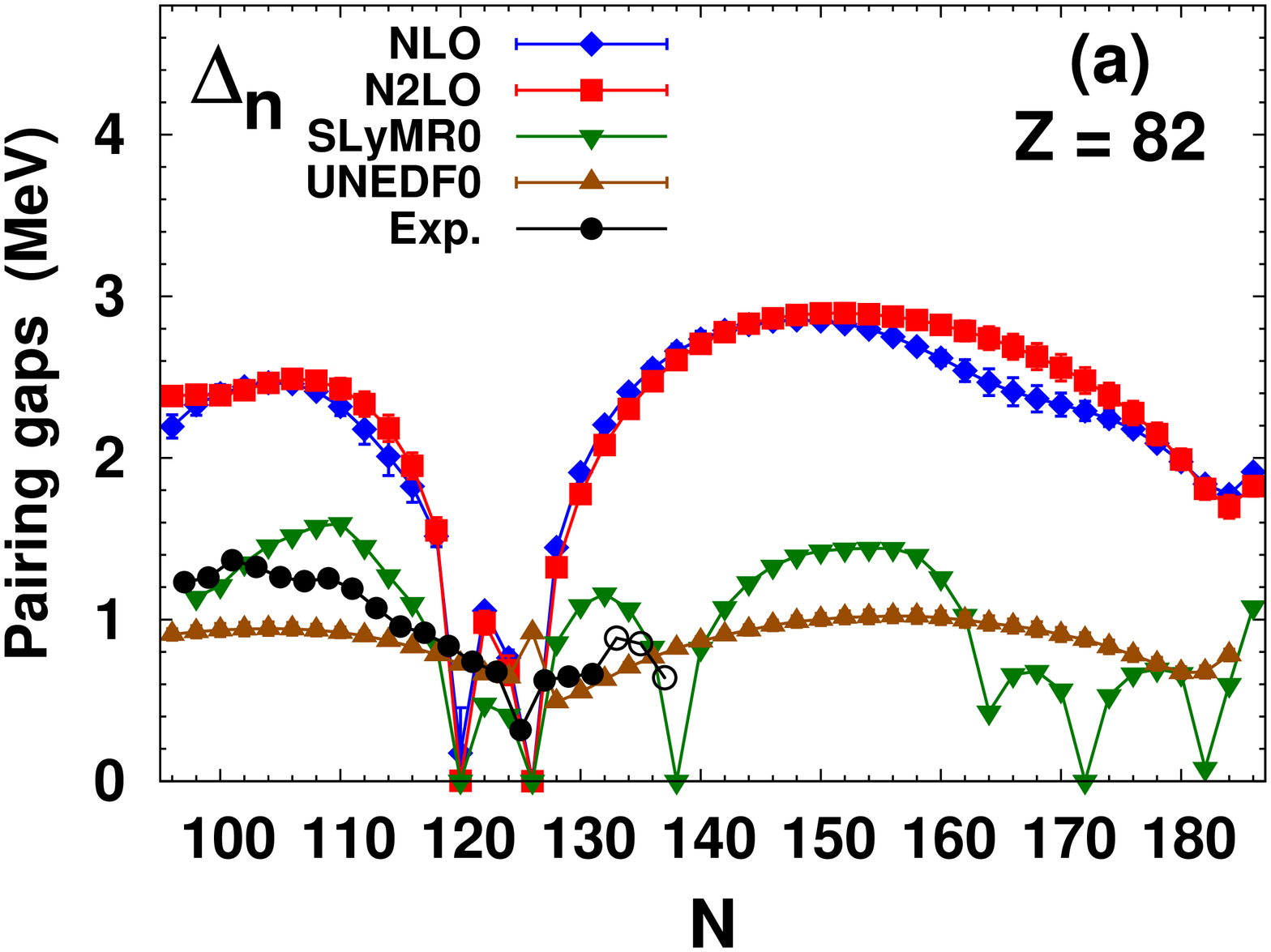} &
\includegraphics[height=0.59\mywidth,angle=270,viewport=40 175 525 890]{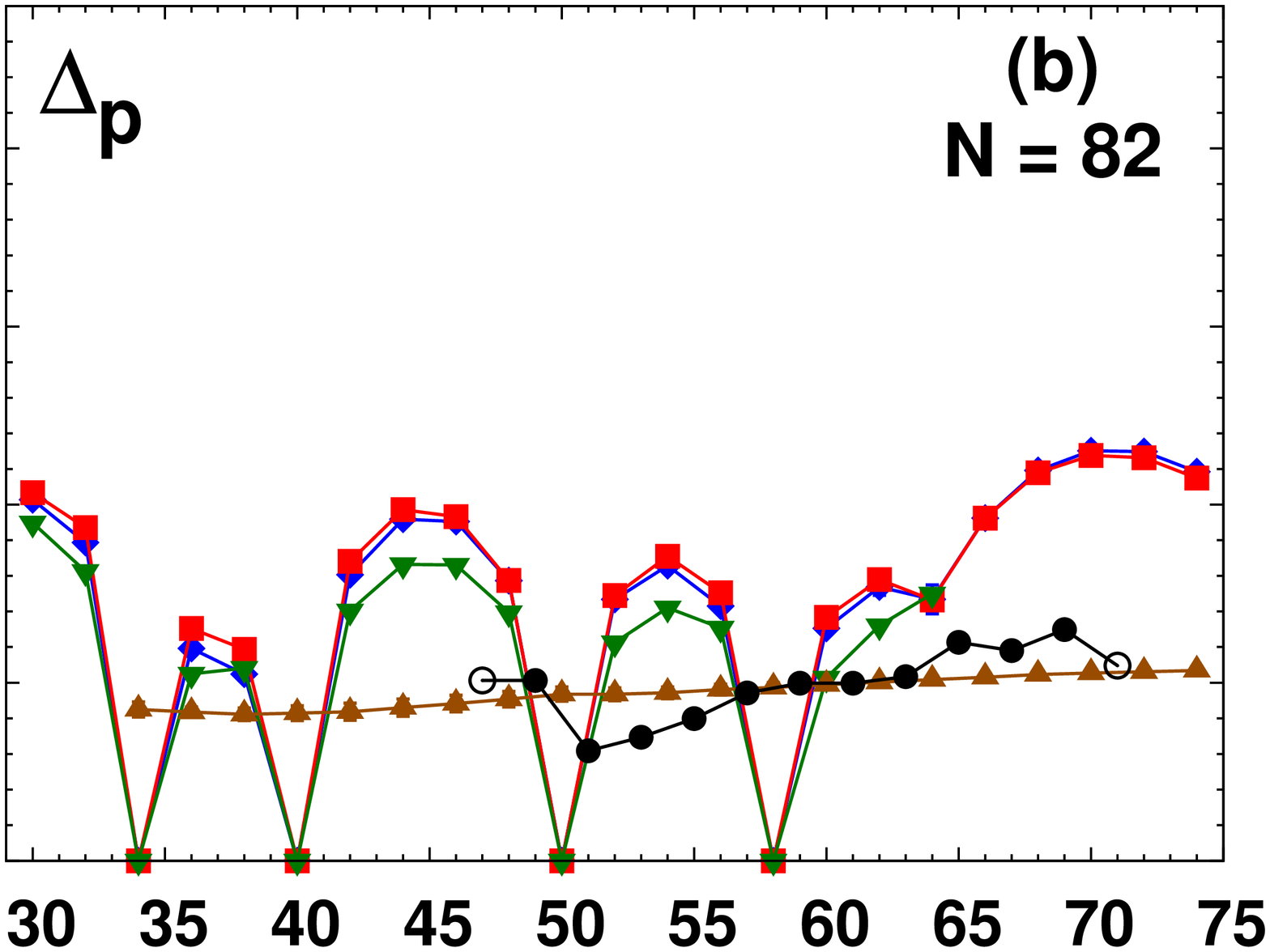} \\[-0.05\mywidth]
 &
\includegraphics[height=0.59\mywidth,angle=270,viewport=40 175 525 890]{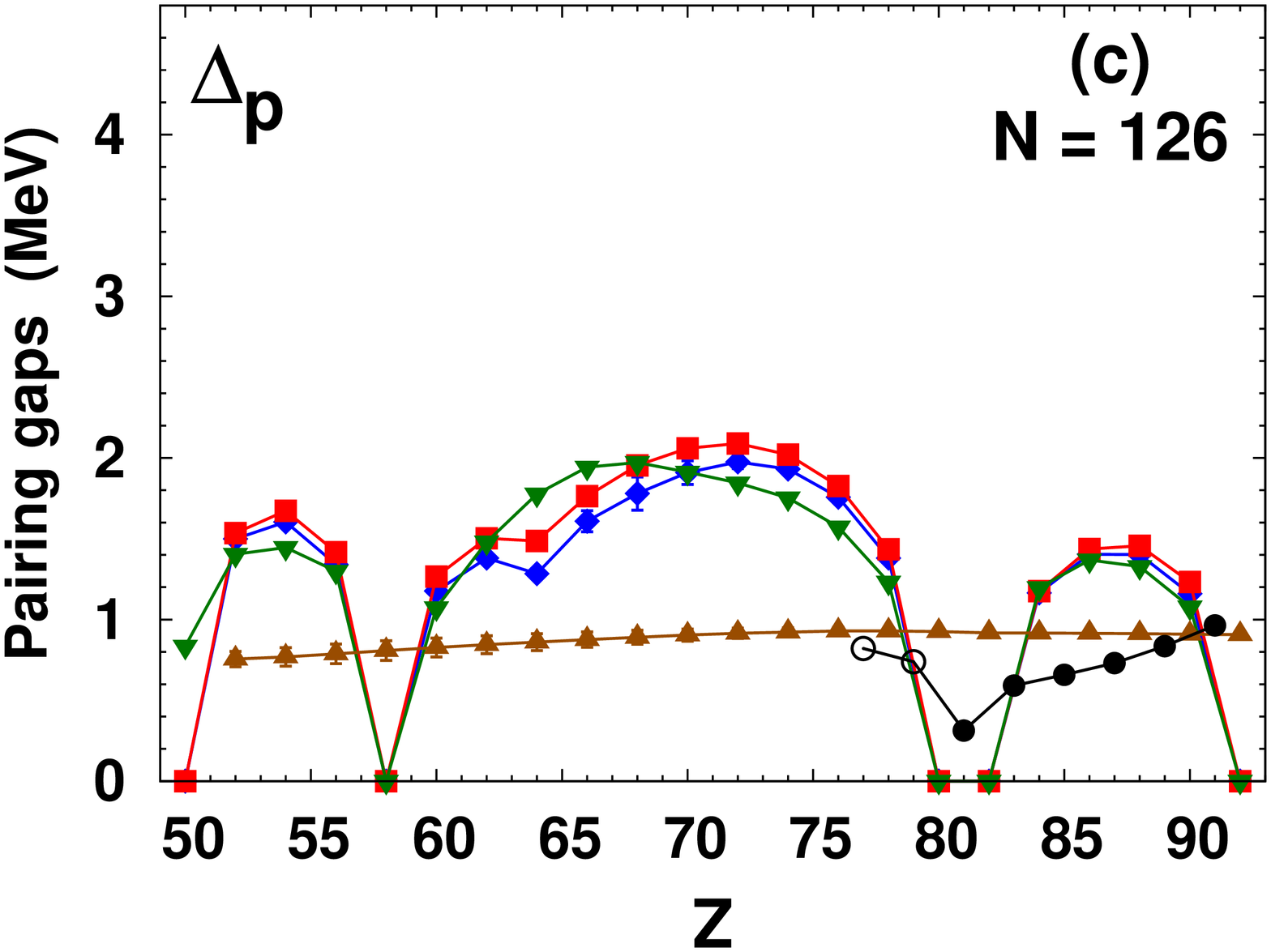}
\end{tabular}
\end{center}
\caption{Same as in Fig.~\ref{fig:fig3} but for the semi-magic nuclei with
$Z,N=82$ and $N=126$.}
\label{fig:fig4}
\end{figure}

In Figs.~\ref{fig:fig3} and~\ref{fig:fig4}, we show neutron and
proton pairing gaps calculated as pairing fields averaged with
density matrices,\cite{(Sto03)} and for the LN method (UNEDF0),
corrected by adding the corresponding $\lambda_2$ LN
parameters.\cite{(Sto03)} As discussed in Ref.\citenum{(Ben17)}, for
the NLO and N$^2$LO EDFs, pairing correlations were adjusted to
values largely overestimating experimental data. This feature is
clearly visible in the figures, and should certainly be improved upon
in future planned adjustment of parameters. We also see that the HFB
results shown for the NLO, N$^2$LO, and SLyMR0 EDFs exhibit
unphysical breaking of pairing at doubly magic gaps, which is a
feature related to the lack of particle-number restoration, and thus
is absent in the HFB+LN results shown for UNEDF0.

\section{Conclusions}\label{sec:Conclusions}

In this article we presented the binding-energy residuals and average
pairing gaps obtained for semi-magic nuclei using two recently
adjusted finite-range pseudopotentials at NLO and
N$^2$LO, as well as the EDF UNEDF0 and
zero-range pseudopotential SLyMR0. For all cases but SLyMR0, the
propagated statistical errors of observables were calculated.

For the set of nuclei considered here, and for finite-range pseudopotentials, the average deviations between
experimental and calculated binding energies
are larger than those obtained for EDF UNEDF0, but they are significantly smaller than the ones
obtained for the zero-range pseudopotential SLyMR0. For the NLO and N$^2$LO EDFs, the typical
arches that appear in the binding-energy residuals between major
shells might be
due to their very low effective masses (close to 0.4). However, the fact that
similar arches do not appear for SLyMR0, which has an effective mass
of 0.47, questions this conjecture.

Based on the results obtained for binding-energy residuals and
average pairing gaps, there was no clear improvement when going from
the finite-range pseudopotential at NLO to the one at N$^2$LO. This
does not necessarily mean that the additional degrees of freedom
introduced at N$^2$LO are not relevant, but most likely reflects the
fact that the penalty function based on spherical doubly magic nuclei
did not allow for properly constraining them.

The next step in the development of this family of pseudopotentials
will consist in increasing the effective mass. The obvious way to do
this is to introduce three-body terms in the pseudopotential. The
present computational ressources restrict this extension to the
introduction of zero-range three-body terms in the spirit of the work
of Onishi and Negele~\cite{ONISHI1978336} and will require the use of
a cut-off to prevent the divergence of the energy. The work along
these lines is in progress.

\section*{Acknowledgments}

This work was supported by the Academy of Finland, the University of
Jyv\"askyl\"a within the FIDIPRO program, by the CNRS/IN2P3
through PICS No. 6949.


\bibliographystyle{ws-procs9x6} 

\begin{thebibliography}{10}

\bibitem{(Car08)}
B.~G. Carlsson, J.~Dobaczewski and M.~Kortelainen, Local nuclear energy density
  functional at next-to-next-to-next-to-leading order, {\em Phys. Rev. C} {\bf
  78}, p. 044326  (2008).

\bibitem{(Zal08)}
M.~Zalewski, J.~Dobaczewski, W.~Satu\l{}a and T.~R. Werner, Spin-orbit and
  tensor mean-field effects on spin-orbit splitting including self-consistent
  core polarizations, {\em Phys. Rev. C} {\bf 77}, p. 024316 (Feb 2008).

\bibitem{(Rai11)}
F.~Raimondi, B.~G. Carlsson and J.~Dobaczewski, {Effective} pseudopotential for
  energy density functionals with higher-order derivatives, {\em Phys. Rev. C}
  {\bf 83}, p. 054311  (2011).

\bibitem{(Dob12)}
J.~Dobaczewski, K.~Bennaceur and F.~Raimondi, Effective theory for low-energy
  nuclear energy density functionals, {\em J. Phys. G} {\bf 39}, p. 125103
  (2012).

\bibitem{(Rai14)}
F.~Raimondi, K.~Bennaceur and J.~Dobaczewski, Nonlocal energy density
  functionals for low-energy nuclear structure, {\em J. Phys. G: Nucl. Part.
  Phys.} {\bf 41}, p. 055112  (2014).

\bibitem{(Ben14a)}
K.~Bennaceur, J.~Dobaczewski and F.~Raimondi, Extended skyrme pseudopotential
  deduced from infinite nuclear matter properties, {\em EPJ Web of Conf.} {\bf
  66}, p. 02031  (2014).

\bibitem{(Sad13)}
J.~Sadoudi, M.~Bender, K.~Bennaceur, D.~Davesne, R.~Jodon and T.~Duguet, Skyrme
  pseudo-potential-based EDF parametrization for spuriosity-free MR EDF
  calculations, {\em Phys. Scr.} {\bf T154}, p. 014013  (2013).

\bibitem{(Sad13b)}
J.~Sadoudi, T.~Duguet, J.~Meyer and M.~Bender, {Skyrme} functional from a
  three-body pseudopotential of second order in gradients: {Formalism} for
  central terms, {\em Phys. Rev. C} {\bf 88}, p. 064326  (2013).

\bibitem{(Dav15)}
D.~Davesne, J.~Navarro, P.~Becker, R.~Jodon, J.~Meyer and A.~Pastore, Extended
  skyrme pseudopotential deduced from infinite nuclear matter properties, {\em
  Phys. Rev. C} {\bf 91}, p. 064303 (Jun 2015).

\bibitem{(Ben17)}
K.~Bennaceur, A.~Idini, J.~Dobaczewski, P.~Dobaczewski, M.~Kortelainen and
  F.~Raimondi, arXiv:1611.09311.

\bibitem{(Kor10b)}
M.~Kortelainen, T.~Lesinski, J.~Mor\'e, W.~Nazarewicz, J.~Sarich, N.~Schunck,
  M.~V. Stoitsov and S.~Wild, Nuclear energy density optimization, {\em Phys.
  Rev. C} {\bf 82}, p. 024313 (Aug 2010).

\bibitem{(Sto03)}
M.~V. Stoitsov, J.~Dobaczewski, W.~Nazarewicz, S.~Pittel and D.~J. Dean,
  {Systematic} study of deformed nuclei at the drip lines and beyond, {\em
  Phys. Rev. C} {\bf 68}, p. 054312  (2003).

\bibitem{[Ben17a]}
{Bennaceur K. {\it et al.\/} 2017, to be submitted to Computer Physics
  Communications}.

\bibitem{(Hoo72)}
R.~H. Hooverman, A technique for numerical solution of the Schroedinger
  equation with non-local potentials, {\em Nuclear Physics A} {\bf 189}, 155
  (1972).

\bibitem{(lenteur)}
K.~Bennaceur, {\textsc{lenteur}} {H}{F}{B} code unpublished.

\bibitem{(Car10b)}
B.~Carlsson, J.~Dobaczewski, J.~Toivanen and P.~Vesel\'{y}, {Solution} of
  self-consistent equations for the {N3LO} nuclear energy density functional in
  spherical symmetry. {The} program {\textsc{hosphe}} (v1.02), {\em Comput. Phys. Comm.}
  {\bf 181}, p. 1641  (2010).

\bibitem{[Car13b]}
{B. G. Carlsson, J. Toivanen, P. Vesel\'y, and Y. Gao, to be published}.

\bibitem{(Gao13)}
Y.~Gao, J.~Dobaczewski, M.~Kortelainen, J.~Toivanen and D.~Tarpanov,
  Propagation of uncertainties in the skyrme energy-density-functional model,
  {\em Phys. Rev. C} {\bf 87}, p. 034324  (2013).

\bibitem{(Wan12)}
M.~Wang, G.~Audi, A.~H. Wapstra, F.~G. Kondev, M.~MacCormick, X.~Xu and
  B.~Pfeiffer, The AME2012 atomic mass evaluation (ii). tables, graphs and
  references, {\em Chin. Phys. C} {\bf 36}, 1603  (2012).

\bibitem{(Dob14)}
J.~Dobaczewski, W.~Nazarewicz and P.-G. Reinhard, {Error} {Estimates} of
  {Theoretical} {Models:} a {Guide}, {\em J. Phys. G: Nucl. Part. Phys.} {\bf
  41}, p. 074001  (2014).

\bibitem{(Sat98)}
W.~Satu\l{}a, J.~Dobaczewski and W.~Nazarewicz, Odd-even staggering of nuclear
  masses: Pairing or shape effect?, {\em Phys. Rev. Lett.} {\bf 81}, 3599
  (1998).

\bibitem{ONISHI1978336}
N.~Onishi and J.~Negele, Two-body and three-body effective interactions in
  nuclei, {\em Nuclear Physics A} {\bf 301}, 336   (1978).

\end{thebibliography}

\end{document}